\documentstyle[aas2pp4,11pt,epsf]{article}
\begin{document}

\title{Fragmentation of Molecular Clouds: The Initial Phase of a Stellar Cluster}
\author{Ralf S.~Klessen,  Andreas Burkert and Matthew R.~Bate}
\affil{Max-Planck-Institut f{\"u}r Astronomie, K{\"o}nigstuhl 17,
69117 Heidelberg, Germany}

\begin{abstract}
The isothermal gravitational collapse and fragmentation of a region
within a molecular
cloud and the subsequent formation of a protostellar cluster is
investigated numerically. The clump mass spectrum which forms during
the fragmentation phase can be well approximated by a power law
distribution $dN/dM \propto M^{-1.5}$. In contrast, the mass spectrum
of protostellar cores that form in the centers of Jeans-unstable clumps
and evolve through accretion and $N$-body interactions is described by
a log-normal distribution with a width that is in excellent agreement
with observations of multiple stellar systems. 
\end{abstract}

\keywords{ISM:kinematics and dynamics --- ISM:structure ---
methods:numerical --- stars:formation}

\section{Introduction}
\label{sec:intro}

Understanding the processes leading to the formation of stars is one of the
fundamental challenges in astronomy and astrophysics.  With the advent of new
observational techniques and instruments, especially in the IR and radio
wavebands, a vast amount of astronomical data about star forming regions has
been accumulated. However, on the theoretical side not much progress has yet
been made. Analytical models of the star formation process are restricted to
describing the collapse of isolated, idealized objects (for an overview see
Whitworth \& Summers 1985). Much the same applies to numerical studies (e.g.
Bonnell \& Bastien 1993, Boss 1997, Burkert et al.~1996, 1997, Nakajima \& Hanawa
1996). Star formation is a complex self-gravitating, magneto-hydrodynamical
problem, which includes the effects of heating and cooling, and feedback
pro\-cesses from newly formed stars. Furthermore, it is influenced by the
galactic environment. Taking into account all these processes with
high spatial resolution exceeds by far present computational capabilities.

Previous numerical simulations of the collapse and fragmentation of
molecular cloud regions have shown that a large number of condensed
objects can indeed form on a dynamical timescale as a result of
gravitational fragmentation (e.g.~Larson 1978, Monaghan \& Lattanzio
1991, Keto et al.~1991). In these studies, the clouds were treated as
isolated gaseous spheres which collapsed completely onto themselves.
Instead, we study a small region embedded in a large, stable molecular
cloud complex where only the overdense regions are able to contract
due to self gravity.  We assume the molecular cloud is supported on
large scales by turbulence and/or other processes.  Previous numerical
models were also strongly constrained by numerical resolution.  Larson
(1978), for example, used just 150 particles in an SPH-like
simulation.  Whitworth et al. (1995) and Turner et al. (1995) were the
first who addressed star formation on larger scales in detail using
high-resolution numerical models. However, they studied a different
problem: fragmentation and star formation in the shocked interface of
colliding molecular clumps.  While clump-clump interactions are
expected to be abundant in molecular clouds, the rapid formation of a
whole star cluster requires gravitational collapse on a larger scale
which contains many clumps and dense filaments.

In this letter, we extend previous studies of the collapse of isolated
objects to the regime of the isothermal collapse and fragmentation of
a gravitationally unstable {\em region} embedded in the interior of a
molecular cloud. We present a high-resolution numerical model of the
dynamical evolution and follow the fragmentation into dense
protostellar cores.  The temperature and the density is chosen such
that the region is highly gravitationally unstable and forms a
hierarchically-structured protostellar cluster.  The results of this
study, i.e. the properties of the dense clumps and of the newly formed
protostellar cores are compared with observations.

\section{Numerical Technique}
\label{sec:num-technique}

To follow the time evolution of the system, we use smoothed particle
hydrodynamics (SPH: for a review see Monaghan 1992) which is intrinsically
Lagrangian and can resolve very high density contrasts. The code is based on a
version originally developed by Benz (1990).  We adopt a standard description
of artificial viscosity (Monaghan \& Gingold 1983) with the parameters
$\alpha_v = 1$ and $\beta_v = 2$. The smoothing lengths are variable in space
and time such that the number of neighbors for each particle remains at
approximately fifty. The system is integrated in time using a second order
Runge-Kutta-Fehlberg scheme, allowing individual timesteps for each particle
(Bate et al.~1995). Once a highly-condensed object has formed in the center of
a collapsing cloud fragment and has passed beyond a certain density, we
substitute it by a `sink' particle which then continues to accrete material
from its infalling gaseous envelope (Bate et al. 1995). By doing so we prevent
the code time stepping from becoming prohibitively small. This procedure
implies that we cannot describe the evolution of gas inside such a sink
particle.  However, at some stage of the gravitational collapse the SPH
resolution limit (Bate \& Burkert 1997) would be reached in the fragment
anyway.  For a detailed description of the physical processes inside a
protostellar core, i.e. its further collapse and fragmentation, a new
simulation just concentrating on this single object with the appropriate
initial conditions taken from the larger scale simulation would be
necessary (Burkert at al.~1998).

To achieve high computational speed, we have combined SPH with the
special purpose hardware device GRAPE (Sugimoto et al.~1990, Ebi\-suzaki
et al.~1993), following the implementation described by Umemura et
al.~(1993) and in greater detail by Steinmetz (1996).  Since we wish
to describe a region in the interior of a globally-stable molecular
cloud we have to prevent global collapse. Therefore, we use periodic
boundaries, applying the Ewald (1921) method in an PM-like scheme
(Klessen 1997). In this letter we present a simulation with $500\;000$
SPH particles.

\section{Initial Conditions for Cloud Fragmentation}
\label{sec:initial-conditions}
The structure of molecular clouds is very complex, consisting of a
hierarchy of clumps and filaments on all scales (for a review see
Blitz 1993). Many attempts have been made to identify the clump
structure and derive its properties (Stutzki \& G{\"u}sten 1990,
Williams et al. 1994).  We choose as starting conditions density
fields with Gaussian random fluctuations that follow a power spectrum
$P(k) \propto 1/k^2$, i.e. large scale fluctuations have on average
large amplitudes whereas the amplitudes of short wavelength modes
decay quadratically. The fields are generated by applying the
Zel'dovich (1970) approximation to an originally homogeneous gas
distribution: we compute a hypothetical field of density fluctuations
in Fourier space and solve Poisson's equation to obtain the
corresponding self-consistent velocity field.  These velocities are
then used to advance the particles in one big step $\delta t$. 

\begin{figure*}[th]
\unitlength1.2cm
\begin{picture}(12,9.5)
\put( 1.3, 4.8){\epsfxsize=6.7cm \epsfbox{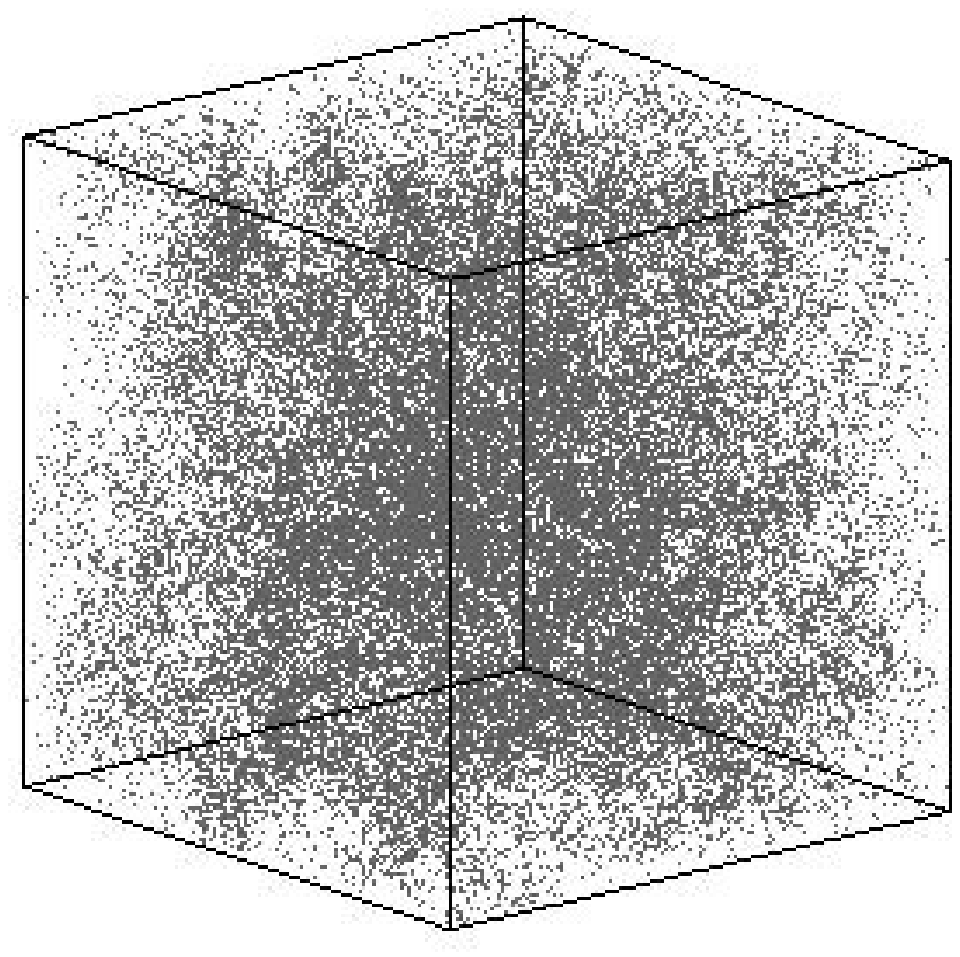}}
\put( 6.5, 4.8){\epsfxsize=6.7cm \epsfbox{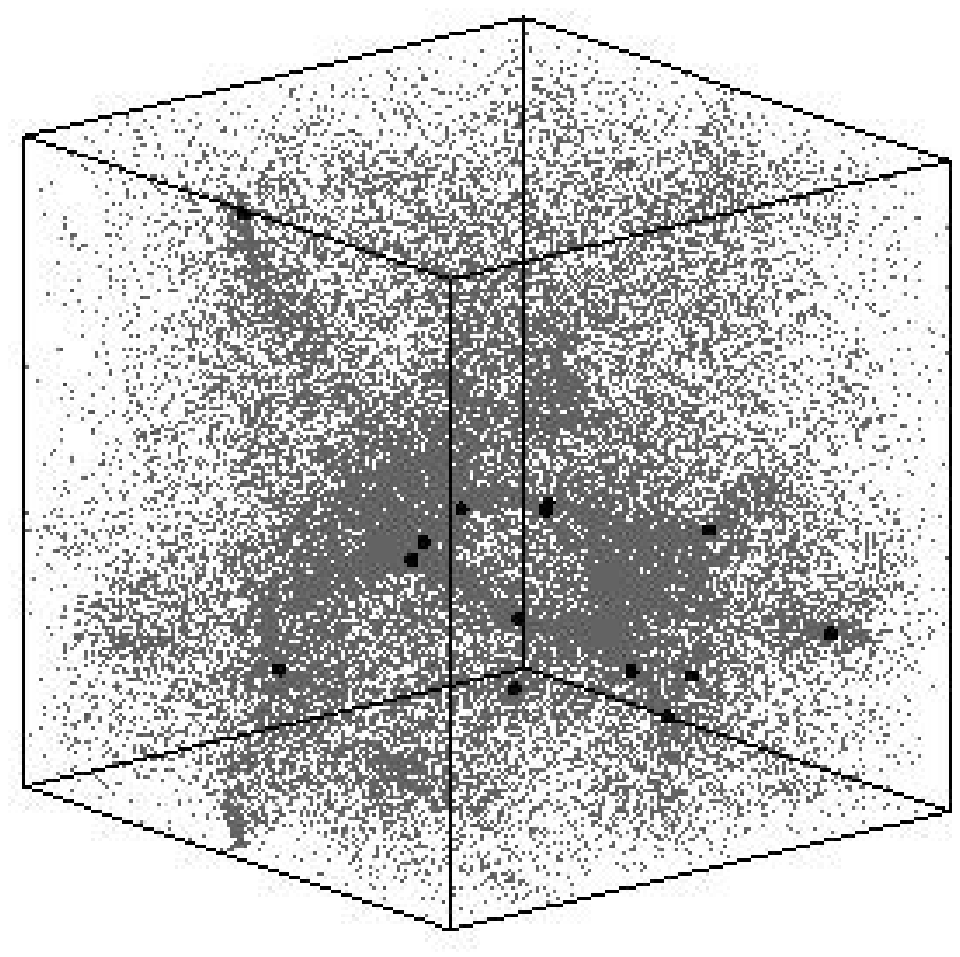}}
\put( 1.3, 0.2){\epsfxsize=6.7cm \epsfbox{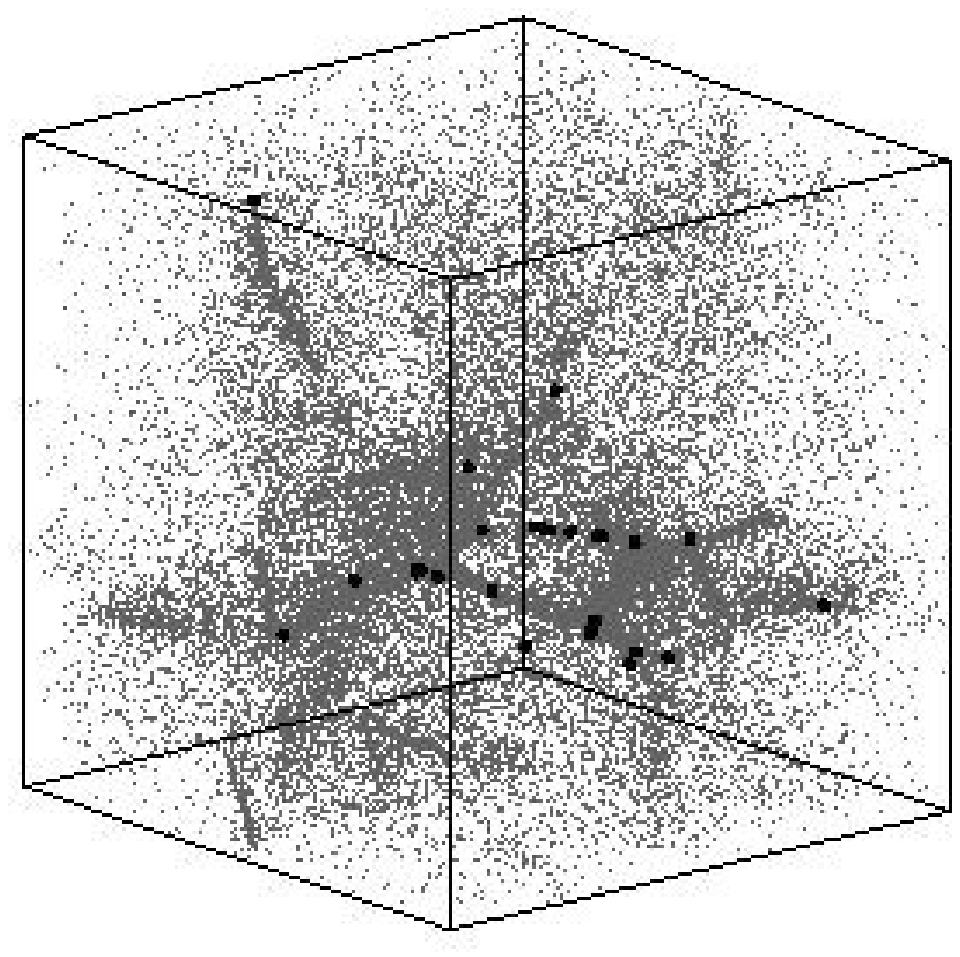}}
\put( 6.5, 0.2){\epsfxsize=6.7cm \epsfbox{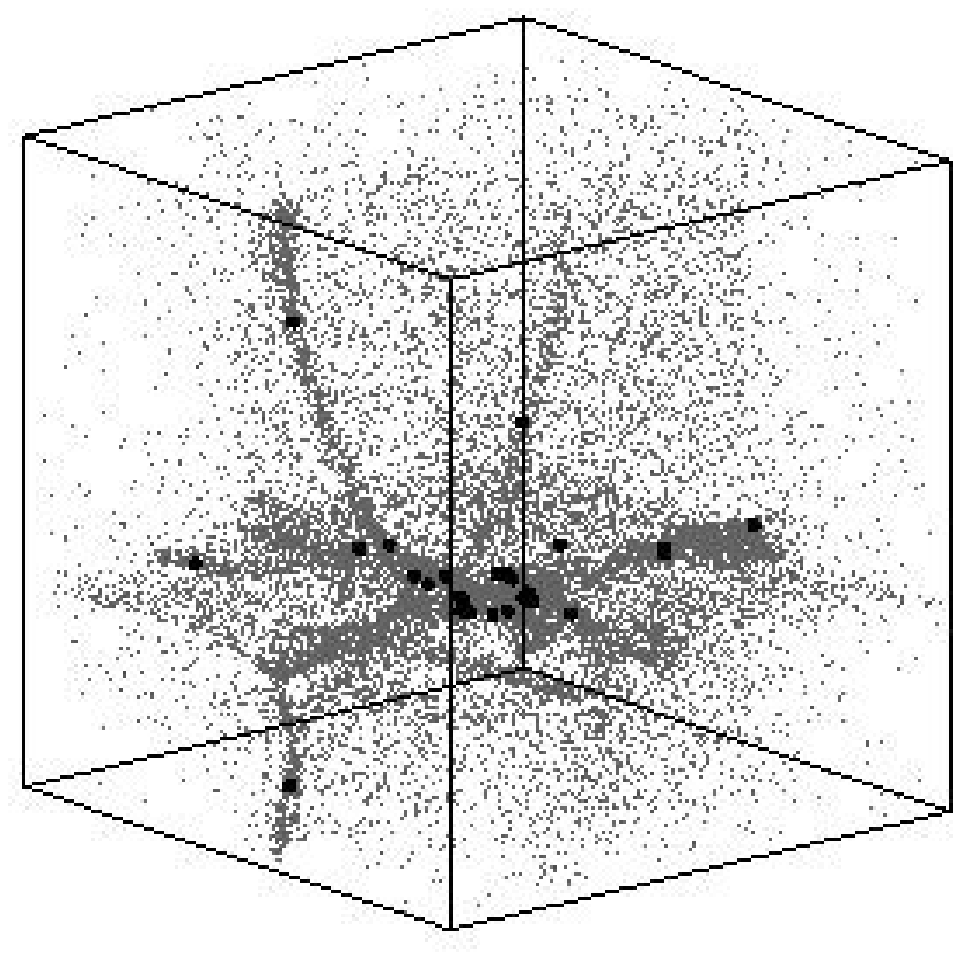}}
\put( 1.8, 5.0){ $t=0.0$}
\put( 7.0, 5.0){ $t=1.6$}
\put( 1.8, 0.4){ $t=2.0$}
\put( 7.0, 0.4){ $t=2.8$}
\end{picture}
\caption{
\label{fig:3D-plots}
Time evolution and fragmentation of a region of 222 Jeans masses in
the interior of a molecular cloud with initial Gaussian density
fluctuations with power law $P(k) \propto 1/k^2$.  Collapse sets in
and soon forms a cluster of highly-condensed cores, which continue to
accrete from the surrounding gas reservoir. At $t=1.6$ about 10\% of
all the gas mass is converted into "protostellar" cores (denoted by
black dots). At $t=2.0$ and $t=2.8$ these values are 30\% and 60\%,
respectively. The initial number of particles used for the SPH
simulation is $500\;\!000$. For legibility only every tenth particle
is plotted.}
\end{figure*}
\section{A Case Study}
\label{sec:case-study}
As a case study, we present the time evolution of a region in the
interior of a molecular cloud containing a total mass of 222 Jeans
masses, determined from the temperature and mean density of the gas.
Figure~\ref{fig:3D-plots} depicts snapshots of the system initially,
and when 10, 30 and 60 per cent of the gas mass has been accreted onto
the protostellar cores. Note that this cube has to be seen
periodically replicated in all directions. Initially, pressure smears
out small-scale features, whereas large-scale fluctuations start to
collapse on themselves, and into filaments and knots.  After $t \approx
0.9$\footnote{Time is measured in dimension-less units with $t=0$
being defined by the start of the SPH simulation. For adequate timing,
the Zel'dovich shift interval $\delta t = 1.5$ has to be added.  As
reference, the free-fall time of the isolated cube would be $\tau_{\rm
ff} = 1.4$.}, the first highly-condensed cores form in the centers of
the most massive and densest Jeans-unstable gas clumps and are
replaced by sink particles.  Soon, clumps of lower initial mass and
density follow, altogether creating a hierarchically-structured
cluster of accreting protostellar cores.

\subsection{Scaling Properties}
\label{subsec:scaling}
The gas is isothermal. Hence, the calculations are scale free,
depending only on one parameter: the dimensionless temperature
$T\equiv E_{\rm int}/|E_{\rm pot}|$, which is defined as the ratio
between the internal\footnote{In the case of {\em isotropic}
turbulence, the non-thermal (turbulent) contributions can also be
accounted for in this expression, $E_{\rm int} = E_{\rm therm} +
E_{\rm turb}$.}  and gravitational energy of the gas. The model can
thus be applied to star-forming regions with different physical
properties.  In the case of a dark cloud with mean density
$n(H_2)\approx100\,$cm$^{-3}$ and a temperature $T\approx10\,$K like
Taurus-Auriga, the computation corresponds to a cube of length
10$\,$pc and a total mass of $6\,300\,$M$_{\odot}$ (with the Jeans
mass being $M_{\rm J} = 28\,$M$_{\odot}$).  The dimensionless time
unit corresponds to $2.2 \times 10^6\,$yrs. For a high-density
star-forming region like Orion with $n(H_2) \approx 10^5\,$cm$^{-3}$
these values scale to $0.32\,$pc and $200\,$M$_{\odot}$, respectively,
again assuming $T\approx 10\,$K. The Jeans mass is $0.9\,$M$_{\odot}$
and the time scale is $6.9\times10^4\,$yrs.

\subsection{The Importance of Dynamical Interaction and Competitive Accretion}

The location and  time at which protostellar cores form, is
determined by the dynamical evolution of their parental gas
clouds. Besides collapsing individually, clumps stream towards a common
center of attraction where they merge with each other or undergo
further fragmentation. The formation of dense cores in the centers of
clumps depends strongly on the relation between the timescales for
individual collapse, merging and sub-fragmentation.  Individual clumps
may become Jeans unstable and start to collapse to form a condensed
core in their centers. While collapsing, these clumps also interact
with each other. When clumps  merge, the larger new clump
continues to collapse, but contains now a {\em multiple} system of
cores in its center.  Now sharing a common environment, these cores
compete for the limited reservoir of gas in their surrounding (see
e.g. Price \& Podsiadlowski 1995, Bonnell et al.~1997).  Furthermore,
the ``protostellar'' cores interact gravitationally with
each other.  As in dense stellar clusters, close encounters lead to
the formation of unstable triple or higher order systems and alter the
orbital parameters of the cluster members. As a result, a considerable
fraction of ``protostellar'' cores get expelled from their parental
clump.  Suddenly bereft of the massive gas inflow from their
collapsing surrounding, they effectively stop accreting and their
final mass is determined. Ejected objects can travel quite far and
resemble the weak line T Tauri stars found via X-ray observation
in the vicinities of star-forming molecular clouds (e.g.~Neuh{\"a}user et
al.~1995, Wichmann et al.~1997).

\begin{figure}[t]
\epsfxsize=8.0cm \epsfbox{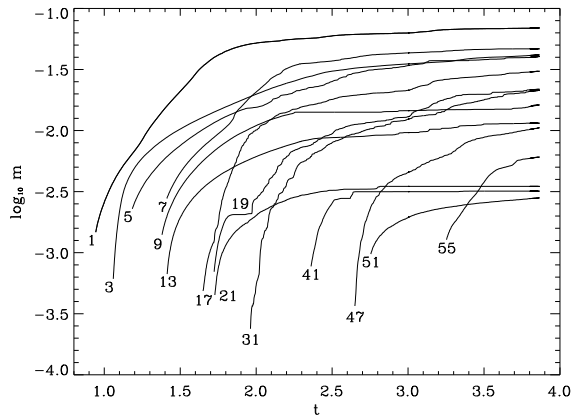}
\caption{
\label{fig:accr-hist}
Mass growth history for 14 selected protostellar cores selected from a
total of 55. The numbers reflect the order of their formation.
}
\end{figure}
In Fig.~\ref{fig:accr-hist}, we plot the accretion history for 14
representative protostellar cores in the simulation.  The objects are
numbered according to their time of formation. The figure illustrates
the following trends: (a) The cores which form first tend to have the
largest final masses. They emerge from the initial clumps with the
highest densities. The collapse of these large initial clumps
contributes a considerable fraction of their total mass.  (b) Matter
that contracts into dense cores at later times (say
$t\:\lower0.6ex\hbox{$\stackrel{\textstyle >}{\sim}$}\:2$) has already
undergone considerable dynamical evolution. Small initial clumps
stream towards each other along filaments. At the intersections of
filaments they merge and may undergo rapid collapse, when enough mass
is accumulated. (c) Once dense cores have formed, they evolve due to
accretion, competing for gas from the surrounding reservoir and
interacting dynamically, as described above.  For example, at
$t\approx 1.8$, core \#19 is expelled from a dense clump at the
intersection of two massive filaments by a triple interaction with
cores \#1 and \#17. It stops accreting. However, it still is bound to
the gas knot which grows in mass due to continuous infall. It falls
back onto the clump of gas and resumes accreting at
$t\approx2.0$. Cores \#9 and \#41 are also expelled from their
parental clumps but, unlike core \#19, their accretion is terminated
completely.  These dynamical interactions between cores are an
important agent in shaping the mass distribution.

\subsection{Mass Spectrum -- Implications for the IMF}
\begin{figure*}[t]
\unitlength1cm
\begin{picture}(16,6.0)
\put( 1.75, 3.10){\epsfxsize=3.25cm \epsfbox{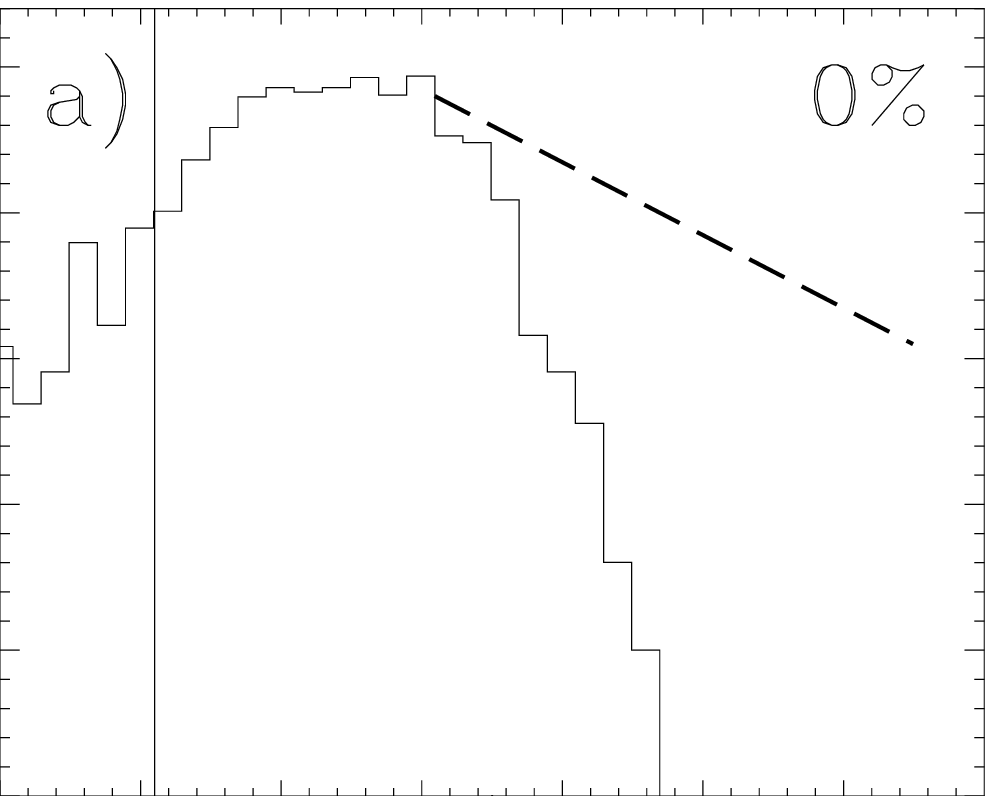}}
\put( 5.00, 3.10){\epsfxsize=3.25cm \epsfbox{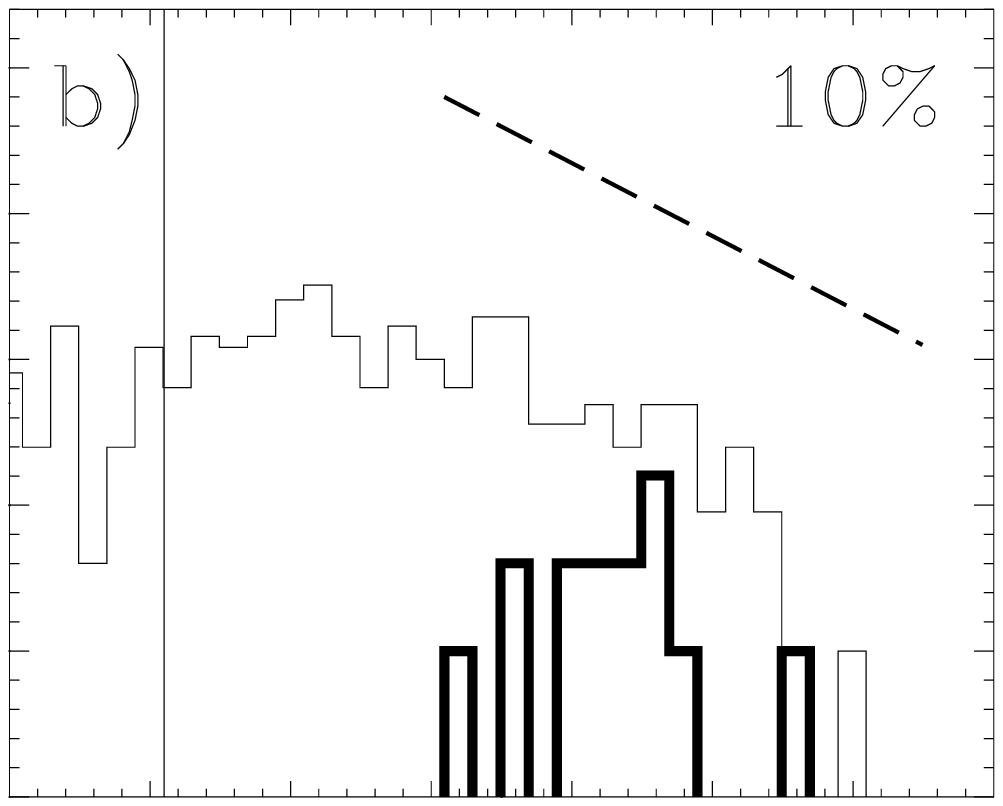}}
\put( 1.75, 0.50){\epsfxsize=3.25cm \epsfbox{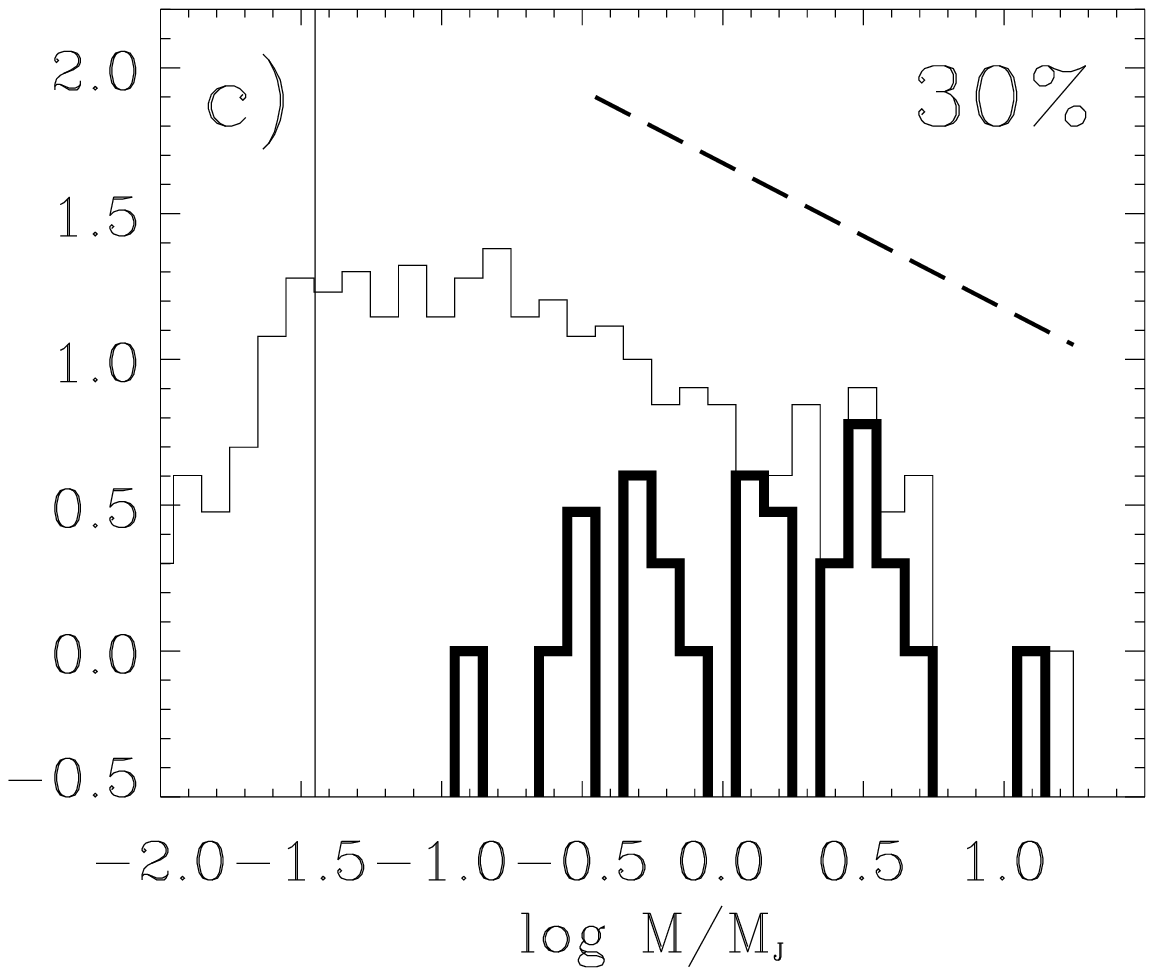}}
\put( 5.00, 0.50){\epsfxsize=3.25cm \epsfbox{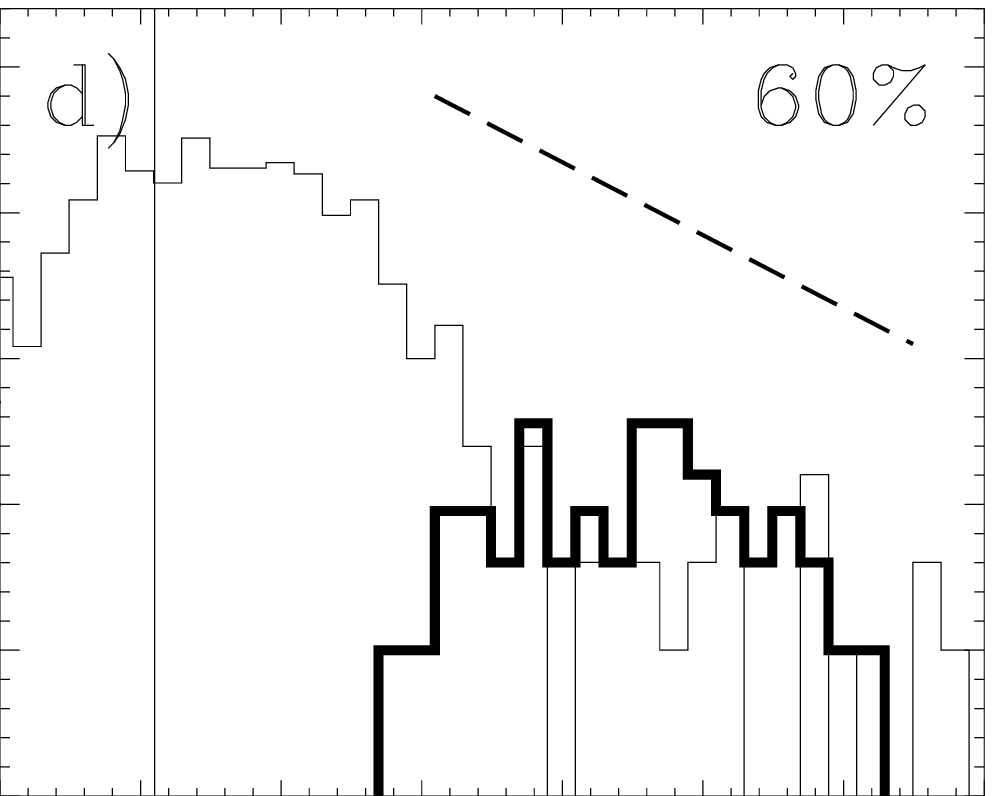}}
\put( 9.75, 0.50){\epsfxsize=6.50cm \epsfbox{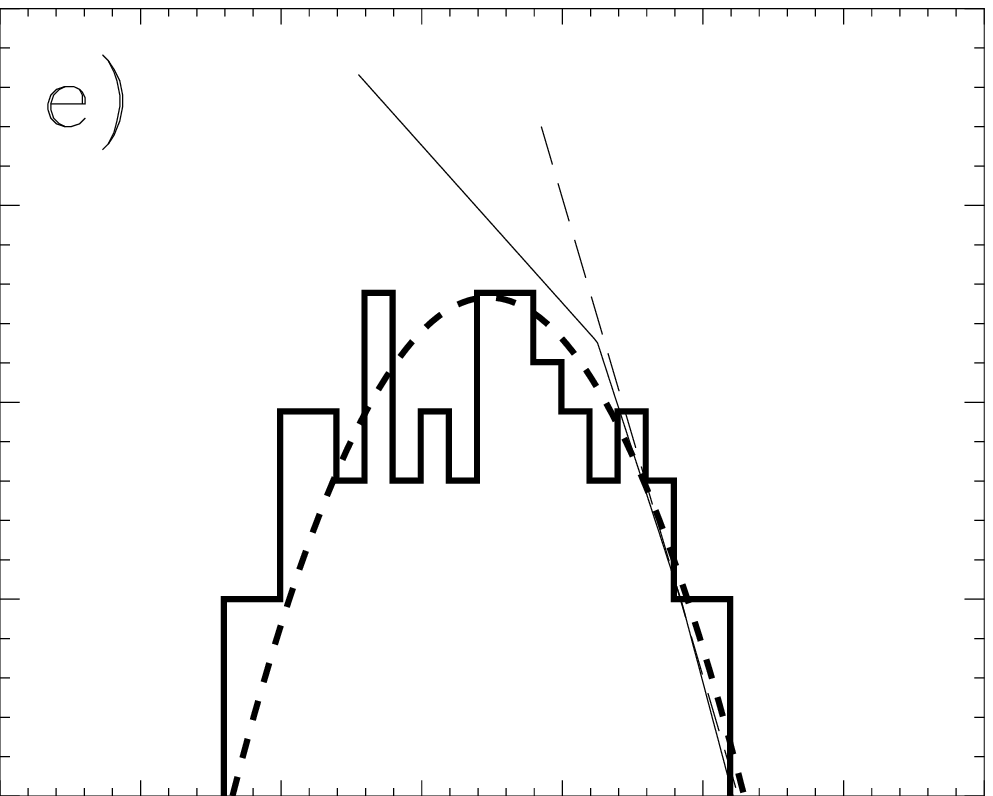}}
\end{picture}
\caption{\label{fig:mass-distr} a) -- d) Mass distribution of gas
clumps (thin lines) and of protostellar cores (thick lines) at times
$t=0.0$, 0.7, 1.3 and 2.0 when 0\%, 10\%, 30\% and 60\% of the total
gas mass is condensed in cores, respectively. The vertical lines
indicate the resolution limit of the simulation with $500\,000$
particles, and the dashed lines illustrate the observed clump mass
spectrum with $dN/dM \propto M^{-1.5}$ (Blitz 1993). e) Comparison of
the final core mass spectrum (thick line) with different
observationally based models for the IMF.  The thick dashed line
denotes the log-normal form for the IMF, uncorrected for binary stars
as proposed by Kroupa et al.~(1990). In order for the peaks of both
distributions to overlap, a core star formation efficiency $\eta$ has
to be assumed.  The agreement in width is remarkable.  The multiple
power-law IMF, corrected for binary stars (Kroupa et al.~1993) is
shown by the thin solid line.  As reference, the thin dashed line
denotes the Salpeter (1955) IMF. Both are scaled to fit at the
high-mass end of the spectrum.  All masses are scaled to the overall
Jeans mass in the system.}
\end{figure*}

Figures~\ref{fig:mass-distr}a -- d describe the mass distribution of
identified gas clumps (thin lines) and of protostellar cores (thick
lines). To identify individual clumps we have developed an algorithm
similar to the method described by Williams et al.~(1994), but based
on the framework of SPH.  As a reference, we also plot the observed
canonical form for the clump mass spectrum, $dN/dM \propto M^{-1.5}$
(Blitz 1993), which has a slope of $-0.5$ when plotting $N$ versus
$M$. Note that our initial condition does not exhibit a clear power
law clump spectrum, but instead consists preferentially of small scale
fluctuations. However, these are quickly damped by pressure forces and
during the subsequent non-linear gravitational collapse a power-law
mass spectrum is formed with a slope that is similar to the observed
clump mass spectrum (Fig.~\ref{fig:mass-distr}b,c).  In all panels,
the vertical lines indicate the SPH resolution limit for $500\,000$
particles (Bate \& Burkert 1997).

A common feature in all our simulations is the broad mass spectrum of
``protostellar'' cores which peaks slightly above the overall {\em
Jeans mass} of the system.  This is somewhat surprising, since the
initial fluctuations span a vast range of masses and peak densities,
and the evolution of each core is heavily influenced by complex
merging and collapse processes. However, in a statistical sense, the
system retains ``knowledge'' of its (initial) average properties. The
present simulations cannot resolve subfragmentation in condensed
cores.  However, detailed simulations show that perturbed cores tend
to break up into multiple systems (e.g.~Burkert et al.~1996, 1997).
Here, we can only determine the mass function of multiple systems
without breaking them down into a mass function of single stars.  Our
simulations predict an initial mass function with a log-normal
functional form, if the mass of a multiple system that forms within a
condensed core is roughly proportional to the core mass, $M_* = \eta
\times M_{core}$. Figure~\ref{fig:mass-distr}e compares the results of
our calculations with the observed IMF for multiple systems (Kroupa et
al.~1990).  The maximum of the observed IMF is located at
$0.23\,$M$_{\odot}$.  The calculated core mass distribution peaks at
$\sim 2\,M_{\rm J}$.  In Section~\ref{subsec:scaling}, we found that
when we scale the simulation to the conditions in low-density
(e.g.~Taurus) and high-density star-forming regions (like Orion), we
obtain Jeans masses of $28\,$M$_{\odot}$ and $0.9\,$M$_{\odot}$,
respectively. Thus, to reproduce the observed peak in the IMF, we take
$\eta \approx 0.005$ for Taurus and $\eta \approx 0.125$ for Orion,
i.e.~the implied star formation efficiency for a low-density
Taurus-like region is much
lower than for cloud regions of high density like Orion. With these
star formation efficiencies, the agreement between the observed IMF
for multiple sytems (thick dashed line; values from Kroupa et
al.~1990) is excellent. Note that although the peak depends on our
choice of $\eta$, the agreement in the {\em width} of the distribution
does not depend on this scaling.  For comparison, the IMF corrected
for binary stars (Kroupa et a.~1993) is indicated as thin solid line,
together with the mass function from Salpeter (1955) as thin dashed
line.

\section{Summary and Discussion}
\label{sec:summary}
 
Since collapse and fragmentation in molecular clouds is an extremely
complex and dynamical process, many authors have sought to understand 
the stellar initial mass function as resulting from a sequence 
of statistical events which may naturally lead 
to a log-normal IMF (see e.g.~Zinnecker 1984, Adams
\& Fatuzzo 1996; also Price \& Podsiadlowski 1995, Murray \& Lin 1996,
Elemegreen 1997).  However, using numerical simulations, it is possible
to identify underlying processes which may contribute to the form of
the stellar initial mass function.  In the calculations presented here,
we find several trends.  The ``protostellar'' cores that form first 
are generally formed in the clumps with the highest initial density,
and tend to have the highest final masses.  Cores that form later,
form from gas that was initially in low-density clumps or distributed
gas which converged to form a higher-density clump before quickly
collapsing.  Overlaid on these general trends, dynamical interactions
between individual cores can act to terminate accretion on to a core by
ejecting it from a clump, thus setting its final mass.  The excellent 
agreement between the numerically-calculated mass function and the
observed IMF for multiple stellar systems (Kroupa et al.~1990) strongly
suggests that these gravitational fragmentation and accretion processes
dominate the origin of stellar masses.  In a subsequent paper, the results
from calculations spanning a larger range of the parameter space 
relevant for molecular clouds shall be discussed in detail.

\end{document}